# Topologically protected elastic waves in one-dimensional phononic crystals of continuous media

Ingi Kim[1]*, Satoshi Iwamoto[1,2], and Yasuhiko Arakawa[1,2]

[1] *Institute of Industrial Science, University of Tokyo, Meguro, Tokyo 153-8505, Japan*
[2] *Institute for Nano Quantum Information Electronics, Meguro, Tokyo 153-8505, Japan*
*E-mail: kim-ingi@iis.u-tokyo.ac.jp

## Abstract

We report the design of silica-based 1D phononic crystals (PnCs) with topologically distinct complete phononic bandgaps (PnBGs) and the observation of a topologically protected state of elastic waves at their interface. By choosing different structural parameters of unit cells, two PnCs can possess a common PnBG with different topological nature. At the interface between the two PnCs, a topological interface mode with a quality factor of ~5,650 is observed in the PnBG. Spatial confinement of the interface mode is also confirmed by using photoelastic imaging technique. Such topologically protected elastic states are potentially applicable for constructing novel phononic devices.

Topological phenomena in quantum Hall, quantum spin Hall systems and topological insulators have been extensively studied in condensed matter physics[1,2]. A hallmark of such phenomena is topologically protected edge states which are robust against defects and imperfection. The presence of the topological edge states is due to topological characters of bulk electronic bands, which is called the bulk-edge correspondence[3,4]. Recently, topological concepts have also been extended to bosonic systems including photonic and phononic structures which support the topologically protected states of light[5–16], acoustic[17–24] and mechanical[25–30] waves for various applications. Most of experimental studies in mechanical systems have focused on discrete structures such as coupled pendula[27,28] or granular chains[30]. Although they are useful for describing topological concepts in mechanical systems, continuous solid structures supporting topological elastic waves are highly expected to realize practical high speed phononic applications. In contrast to the intensive theoretical studies of the topological elastic waves[31–34], there is a lack of an experimental demonstration in the continuous structures. One of the main challenges is due to high modal densities of elastic waves in continuous-solid structures, preventing the formation of complete PnBGs with topologically distinct properties. Very recently, experimental demonstration of topological elastic waves using continuous structures has been reported[35,36]. However, the structures have only partial phononic bandgap (PnBG) which can cause loss of the topological elastic states by coupling with other propagation modes. Thus, it remains a challenge to realize topological elastic waves in continuous structures with complete PnBGs. To the best of our knowledge, experimental demonstration of topological elastic waves in continuous structures with complete PnBGs is very limited even in one-dimensional (1D) periodic system due to their high modal densities.

In this report, we report the experimental realization of topological interface state in solid-structured quasi 1D phononic crystals (PnCs)[37]. In 1D periodic systems, topologically protected edge states are zero-dimensional (0D), localized at the interface between two PnCs with topologically distinct bandgap[22–24,30,33]. The structure we designed consists of two different silica 1D PnCs with complete PnBGs which are topologically distinct. The topological interface states are guaranteed to exist at the interface between the two PnCs due to the bulk-edge correspondence principle. The interface state with 5,650 of mechanical $Q$-factor is observed at 202.38 kHz. In addition, spatial confinement of the topologically protected interface state is also confirmed by the photoelastic imaging method. Such high-frequency topologically localized states has not yet been achieved experimentally in both fluidic[22,23] and solid[30] systems based on 1D PnCs. This demonstration in continuous-solid structures is a significant step towards the realization of practical phononic applications utilizing topological concepts.

Figure 1 (a) shows a unit cell made of fused silica with period $D = 16$ mm along with *x*-direction. The unit cell consists of two wider blocks sandwiching an inner block. Here, we use COMSOL Multiphysics, a commercial package based on the finite-element method, for all numerical simulations in this study. In the simulations, the mass density of fused silica 2200 kg/m$^3$, the longitudinal and shear wave speeds in silica $v_l$=5972 m/s and $v_s$=3766 m/s respectively were used. Analogous to the Su–Schrieffer–Heeger (SSH) model for polyacetylene[38], different structural parameters of the unit cell can provide different topological properties of bulk bands, characterized by Zak phase[39]. We designed topologically distinct PnCs consisting of different unit cells by tuning the structural parameters of unit cells: U0 ($W_{in}/D = 0.5, d_{in}/D = 0.685$), U1 ($W_{in}/D = 0.3, d_{in}/D$ =0.59) and U2 ($W_{in}/D = 0.505, d_{in}/D$ =0.73). Phononic band structures for each unit cell are presented in Fig. 1 (b), (c) and (d), respectively. In the case of U0, PnBG is closed as shown in Fig. 1 (b). The accidental degeneracy point near 204 kHz is topological phase transition point where two bands near the point can invert and the Zak phase in the bulk band can changes when structural parameters shift properly from those of U0, which are analogous to the band inversion process in electronic systems[1]. The band structure of U1 and U2 which have different parameters with the case of U0 were calculated as shown in Fig. 1 (c) and (d), respectively. The PnBGs are closed and reopened between the structural parameters of U1 and U2 via those of U0. Figure 1 (e) and (f) show *x*-components of normalized displacement fields of two band-edge states at $k_x = 0$ of U1 corresponding to blue and red circles in Fig. 1(c), respectively. The symmetry of the band-edge state is either odd or even due to inversion symmetry with respect to its central *yz*-plane in the unit cell. In the case of U2, the symmetries of the band-edge states, i.e. odd and even symmetries, are reversed via the band inversion process, corresponding to the switched colors of the circles at the band-edge frequencies in Fig. 1 (d). As a result, Zak phases of the lower (upper) band near the PnBG change from π(0) to 0 (π), which indicates the topological transition. The Zak phase for the *n*th isolated band in elastic systems is given by

$$\varphi_n^{\text{Zak}} = i \int_{-\pi/D}^{\pi/D} \langle U_{n,k} | \partial_k | U_{n,k} \rangle dk \quad (1)$$

where $U_{n,k}$ is the normalized Bloch waves of elastic waves with wavevector *k*. When the unit cell has inversion symmetry, the Zak phase must be 0 or $\pi$[16,22,40,41]. Note that Zak phases of U1 and U2 are gauge dependent values which depend on the choice of origin of the unit cell but the difference between the Zak phases of U1 and U2, which is $\Delta\varphi^{\text{Zak}} = \varphi^{\text{Zak(U1)}} - \varphi^{\text{Zak(U2)}} = \pi$, is uniquely defined[42]. The distinct topological properties ensure the existence of interface states between two PnCs, U1 and U2, predicted from the bulk-edge correspondence. The existence of the interface state is related to the sum of all Zak phases below the gap on either side of the interface but has no dependence on the properties of the higher bands[16,22,40]. All Zak phases of U1 and U2 below the PnBG were calculated and eigenfrequencies of topological interface states in the composite PnC, U1+U2, were found in the topologically distinct PnBG regions (see supplementary data, S1). It is worth noting that only single localized state in each overlapped complete PnBG region exists in our structure even in higher-order PnBGs, such as our target PnBG between 20th and 21th bands. The number of the interface states in each overlapped PnBG is determined by the topological character of the gap. In the present 1D system, the binary nature of the Zak phase gives the number of interface state of 0 or 1[43]. The simultaneous realization of single localized elastic mode in several complete PnBG regions is a unique property contrasting the topologically protected localized states from the localized modes in conventional PnC cavities.

In addition to the configuration of U0, in the structural parameter space, $W_{in}/D$ and $d_{in}/D$, there are other configurations showing the topological transition point at $k_x=0$. Thus, there are various combinations of topologically distinct configurations such as U1 and U2. A pair of two configurations with different structural parameters, U1' ($W_{in}/D = 0.33, d_{in}/D = 0.59$) and U2' ($W_{in}/D = 0.501, d_{in}/D = 0.72$) is an example. Those topological properties are identical with that of U1 and U2, respectively (see supplementary data, S2). Among various possibilities, we used the pair and the structure U1 to experimentally demonstrate the topologically protected localized elastic state in silica 1D PnCs.

Figure 2 (a) and (b) shows phononic band structures of composite PnC, U1+U2' and U1+U1', respectively. The composite PnCs consisting of 10-unit cells for each PnC component were used as a supercell repeated periodically in the *x*-direction for the calculation of the band structures. The green regions in the band structures indicate common PnBGs of the PnC components. In the case of U1+U2', the common PnBG is narrower than that of U1+U1' since the lower band-edge frequency in U2' is higher than the others (see supplementary data, S3). As discussed above, the different topological characteristics between U1 and U2' induced the topological interface states in the common PnBG as localized states indicated by the red lines in Fig. 2(a), while no interface state exists in the case of U1+U1' which components are topologically identical. Figure 2 (c) shows the total displacement field of the interface state in the case of the finite-size composite PnC without the periodic boundaries at the edges, U1+U2' with the number of period $N=4$ for each PnC components. The total displacement field is well localized around the interface owing to the PnBG of U1 and U2'. The corresponding eigenfrequency, 202.97 kHz shows a good agreement with the frequency indicated by the red line in Fig. 2(a).

Samples made of fused silica were fabricated by ultrasonic machining process. Two types of composite PnCs, U1+U2 and U1+U1', were fabricated as topologically non-trivial and trivial junctions, respectively. Figure 3 (a) and (b) show the samples of the composite PnCs, U1+U2' and U1+U1', respectively. Each component with the configuration U1, U2' and U1'of the composite PnCs consists of 4-unit cells, i.e. $N=4$. In order to confirm each PnBG property of the components, the individual PnCs with U1, U1' and U2' consisting of 6-unit cells were also fabricated (See supplementary data, S4).

We measured transmission spectrum of elastic waves through the samples. Two transducers were placed on the left and right side of the samples as a transmitter and a receiver, respectively. Glycerin was used as a couplant providing more efficient elastic transmission between the transducers and the samples. The transmission spectra for samples of PnCs consisting of a single configuration of U1, U1' and U2' agree well with the numerical simulations and clearly show bandgap properties, as predicted from each band structure (See supplementary data, S4). Figure 4 (a) and (b) shows the transmission spectra for the samples of the composite PnCs, U1+U2' and U1+U1' shown in Fig. 3(a) and (b), respectively. In Fig. 4 (a), a peak originating from the topological interface state was clearly observed in the common PnBG of U1 and U2' shown in green region. On the other hand, no localized state was observed in the case of topological trivial junction, U1+U1', as shown in Fig. 4 (b) since the PnBGs of U1 and U1' have topologically identical properties.

Figure 5 shows the measured elastic power near the frequency of the peak in Fig. 4(a). The peak in the transmitted elastic power is located at the center frequency of ~202.38 kHz, which is in very good agreement with the calculated eigenfrequency of the interface state, 202.97 kHz in Fig. 2(c). The response curve is well fitted by a single Lorentzian with a full width at half maximum of ~35.8 Hz, which corresponds to the mechanical $Q$~5650.

The strong spatial confinement of the interface state was also experimentally confirmed by photoelastic imaging method. The interface states induce the birefringence in fused silica structures through the photoelastic effect. This produces time-varied retardation of the light passing through the structure. Since the peak retardation is proportional to the difference between the amplitude of strain components at each position of the incident light, the spatial distribution of peak retardation reflects the strain fields of the topological elastic waves[44,45]. The spatial distribution of the interface mode was achieved by measuring the peak retardations at different incident positions of the light. The peak retardation distribution was also calculated for the comparison with the measurement. (see supplementary data for the detail of the methods, S5). Figure 6(a) shows calculated spatial distribution of peak retardation induced by the topological interface state. The black curve in Fig. 6(b) corresponds to the line-cut plot along the yellow broken line in the Fig. 6(a). Here, the retardation is normalized by the value at the interface $x=0$. The results clearly shows the spatial localization properties of the topological interface states. The elastic field decays rapidly from the interface within the second unit cell of each PnC component

In summary, we have designed silica-based 1D PnC possessing topologically distinct complete PnBGs and experimentally demonstrated a topological interface state of elastic wave in the solid-structured PnC. The topologically distinct complete PnBGs are realized by changing the structural parameters of the PnC unit cell. Based on the design, we fabricated a composite PnCs with the interface between two PnCs which have topologically distinct PnBG and demonstrated a highly-confined topological interface state at ~202 kHz with a mechanical $Q$ factor of ~5,650 through the transmission measurement and the photoelastic imaging. To the best of our knowledge, such high quality factor of the topological interface states has not yet been achieved in other topological acoustic/mechanical systems based on 1D PnCs. In monolithic solid structures we used, the frequency of topological elastic states can be easily increased to above MHz by scaling down the feature size of the structures. Thus, our result would be an important step not only for advancing the fundamental research of topological elastic waves but also for realizing practical high-speed elastic devices using topological concepts.

**Acknowledgments**

The authors would like to acknowledge Yasuhiro Hatsugai of University of Tsukuba, Shun Takahashi of Kyoto Institute of Technology and Yasutomo Ota of University of Tokyo for fruitful discussions. This work was partially supported by MEXT KAKENHI Grant Number JP17J09077, JP15H05700, JP17H06138, JP15H05868 and by the Ono Acoustics Research Fund.

# References

[1] M.Z. Hasan and C.L. Kane, Rev. Mod. Phys. **82**, 3045 (2010).

[2] X.L. Qi and S.C. Zhang, Rev. Mod. Phys. **83**, 1057 (2011).

[3] Y. Hatsugai, Phys. Rev. Lett. **71**, 3697 (1993).

[4] Y. Hatsugai, T. Fukui, and H. Aoki, Phys. Rev. B - Condens. Matter Mater. Phys. **74**, 205414 (2006).

[5] F.D.M. Haldane and S. Raghu, Phys. Rev. Lett. **100**, 013904 (2008).

[6] Z. Wang, Y. Chong, J.D. Joannopoulos, and M. Soljacić, Nature **461**, 772 (2009).

[7] L. Lu, J.D. Joannopoulos, and M. Soljačić, Nat. Photonics **8**, 821 (2014).

[8] M. Hafezi, E.A. Demler, M.D. Lukin, and J.M. Taylor, Nat. Phys. **7**, 907 (2011).

[9] M. Hafezi, S. Mittal, J. Fan, a. Migdall, and J.M. Taylor, Nat. Photonics **7**, 1001 (2013).

[10] M.C. Rechtsman, J.M. Zeuner, Y. Plotnik, Y. Lumer, S. Nolte, M. Segev, and A. Szameit, Nature **496**, 196 (2012).

[11] A.B. Khanikaev, S. Hossein Mousavi, W.-K. Tse, M. Kargarian, A.H. MacDonald, and G. Shvets, Nat. Mater. **12**, 233 (2012).

[12] W.-J. Chen, S.-J. Jiang, X.-D. Chen, B. Zhu, L. Zhou, J.-W. Dong, and C.T. Chan, Nat. Commun. **5**, 5782 (2014).

[13] L. Lu, Z. Wang, D. Ye, L. Ran, F. Liang, J.D. Joannopoulos, and M. Soljačić, Science, **349**, 622 (2015).

[14] K.Y. Bliokh, D. Smirnova, and F. Nori, Science, **348**, 1448 (2015).

[15] L.-H. Wu and X. Hu, Phys. Rev. Lett. **114**, 223901 (2015).

[16] M. Xiao, Z.Q. Zhang, and C.T. Chan, Phys. Rev. X **4**, 021017 (2014).

[17] A.B. Khanikaev, R. Fleury, S.H. Mousavi, and A. Alù, Nat. Commun. **6**, 8260 (2015).

[18] Z. Yang, F. Gao, X. Shi, X. Lin, Z. Gao, Y. Chong, and B. Zhang, Phys. Rev. Lett. **114**, 114301 (2015).

[19] C. He, Z. Li, X. Ni, X.C. Sun, S.Y. Yu, M.H. Lu, X.P. Liu, and Y.F. Chen, Appl. Phys. Lett. **108**, 031904 (2016).

[20] Z.G. Chen and Y. Wu, Phys. Rev. Appl. **5**, 054021 (2016).

[21] X. Ni, C. He, X.C. Sun, X.P. Liu, M.H. Lu, L. Feng, and Y.F. Chen, New J. Phys. **17**, 053016 (2015).

[22] M. Xiao, G. Ma, Z. Yang, P. Sheng, Z.Q. Zhang, and C.T. Chan, Nat. Phys. **11**, 240 (2015).

[23] Y.X. Xiao, G. Ma, Z.Q. Zhang, and C.T. Chan, Phys. Rev. Lett. **118**, 166803 (2017).

[24] Z. Yang, F. Gao, and B. Zhang, Sci. Rep. **6**, 29202 (2016).

[25] E. Prodan and C. Prodan, Phys. Rev. Lett. **103**, 248101 (2009).

[26] T. Kariyado and Y. Hatsugai, Sci. Rep. **5**, 18107 (2016).

[27] L.M. Nash, D. Kleckner, A. Read, V. Vitelli, A.M. Turner, and W.T.M. Irvine, Proc. Natl. Acad. Sci. **112**, 14495 (2015).

[28] R. Susstrunk and S.D. Huber, Science, **349**, 47 (2015).

[29] G. Salerno, T. Ozawa, H.M. Price, and I. Carusotto, Phys. Rev. B **93**, 085105 (2016).

[30] R. Chaunsali, A. Thakkar, E. Kim, P.G. Kevrekidis, and J. Yang, Phys. Rev. Lett. **119**, 024301 (2017).

[31] N. Swinteck, S. Matsuo, K. Runge, J.O. Vasseur, P. Lucas, and P. a. Deymier, J. Appl. Phys. **118**, 063103 (2015).

[32] S.H. Mousavi, A.B. Khanikaev, and Z. Wang, Nat. Commun. **6**, 8682 (2015).

[33] H. Huang, J. Chen, and S. Huo, J. Phys. D. Appl. Phys. **50**, 275102 (2017).

[34] R.K. Pal and M. Ruzzene, New J. Phys. **19**, 025001 (2017).

[35] J. Vila, R.K. Pal, and M. Ruzzene, arXiv:1705.08247

[36] S. Yu, C. He, Z. Wang, F. Liu, X. Sun, Z. Li, M.-H. Lu, X.-P. Liu, and Y.-F. Chen, arXiv:1707.04901

[37] I. Kim, S. Iwamoto, and Y. Arakawa, Proc. Int. Conf. 8th Metamaterials, Photonic Cryst. Plasmon.(META), 2017, p.359.

    We note that there are comprehensive analysis of the results that were not disclosed in the proceeding, especially design of the structures with their topological characters. More comprehensive explanations discussed in our manuscript and supplement data are necessary to clearly demonstrate existence and observation of the topological elastic states in the structures.

[38] W.P. Su, J.R. Schrieffer, and A.J. Heeger, Phys. Rev. Lett. **42**, 1698 (1979).

[39] J. Zak, Phys. Rev. Lett. **62**, 2747 (1989).

[40] J. Zak, Phys. Rev. B **32**, 2218 (1985).

[41] W. Kohn, Phys. Rev. **115**, 809 (1959).

[42] M. Atala, M. Aidelsburger, J.T. Barreiro, D. Abanin, T. Kitagawa, E. Demler, and I. Bloch, Nat. Phys. **9**, 795 (2013).

[43] P. Delplace, D. Ullmo, and G. Montambaux, Phys. Rev. B **84**, 195452 (2011).

[44] I. Kim, S. Iwamoto, and Y. Arakawa, Jpn. J. Appl. Phys. **55**, 08RD02 (2016).

[45] I. Kim, S. Iwamoto, and Y. Arakawa, Proc. 22th Microoptics Conf. (MOC), 2017, G-5

**Figure Captions**

**Fig. 1.** (a) Unit cell of the PnCs used in this study. (b-d) Phononic band structures of U0, U1 and U2, respectively. (b) The accidental degeneracy point at $k_x$=0 is a topological transition point. (c, d) Two bands shown in blue and red lines are inverted and the Zak phase changes. (e, f) The *x*-components of normalized displacement fields with odd and even symmetries of two band-edge states in the case of U1, corresponding to the blue and red circles at $k_x$=0 in (c), respectively. The symmetries of the band-edge states are reversed in the case of U2 via band inversion process, corresponding to the switched colors of the circles at $k_x$=0 in (d). PnBGs marked with different colors show different topological characteristics.

**Fig. 2.** Phononic band structures of composite PnCs and topological interface mode in a composite PnC, U1+U2'. The composite PnCs consisting of U1+U2' and U1+U1' shown in the top in (a) and (b) were used as a supercell for the calculation of each band structure. Each PnC component of the composite PnC consists of 10-unit cells, i.e. the number of period *N*=10 for each. The regions colored in green indicate common PnBGs of the PnC components. Topological interface states shown in the red line appear in the common PnBG due to topological distinct properties between U1 and U2', while no localized state exists in the case of U1+U1'. (c) The total displacement field of the topological interface state in the case of a finite-size composite PnC of U1+U2' with *N*=4 for each PnC component. The deformed shape of the structure is shown in an exaggerated scale.

**Fig. 3.** Pictures of samples made of fused silica. (a) A composite PnC consisting of U1 (*N*=4, left) and U2' (*N*=4, right) which are topologically distinct characteristics. (b) A composite PnC consisting of U1 (*N*=4, left) and U1' (*N*=4, right) which are topological identical properties.

**Fig. 4.** Measured (blue) and calculated (black) transmission spectra in the case of (a) U1+U2' and (b) U1+U1', corresponding to the samples in Fig. 3 (a) and (b), respectively. The topological interface state was observed in an overlapped PnBG between U1 and U2 presented in green region. On the other hand, no interface state appears in (c) due to identical topological properties.

**Fig. 5.** Measured elastic power at around the frequency of the interface state. Black solid squares are the experimental data. The red curve shows the Lorentzian function fitted to each data set. The mechanical *Q*-factor of ~5650 is observed with the resonance frequency, 202.38 kHz.

**Fig. 6.** Spatial localization property of the topological interface states in the composite PnC, U1+U2'. (a) Calculated normalized retardation distribution in the *xz*-plane. The yellow broken line indicates measurement positions in the experiment. (b) Peak retardation induced by the interface mode measured in the middle line parallel to the *x*-direction corresponding to the yellow broken line in (a). The red solid squares are the experimental results. The black curve indicated the calculated result.

# Figures

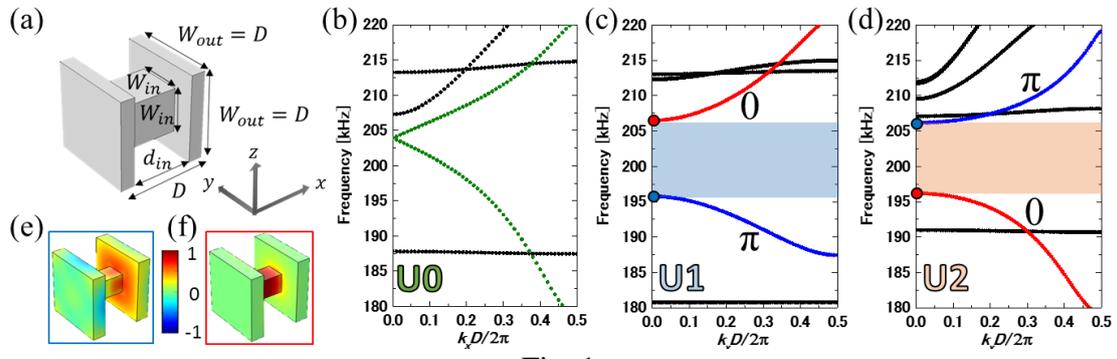

Fig. 1.

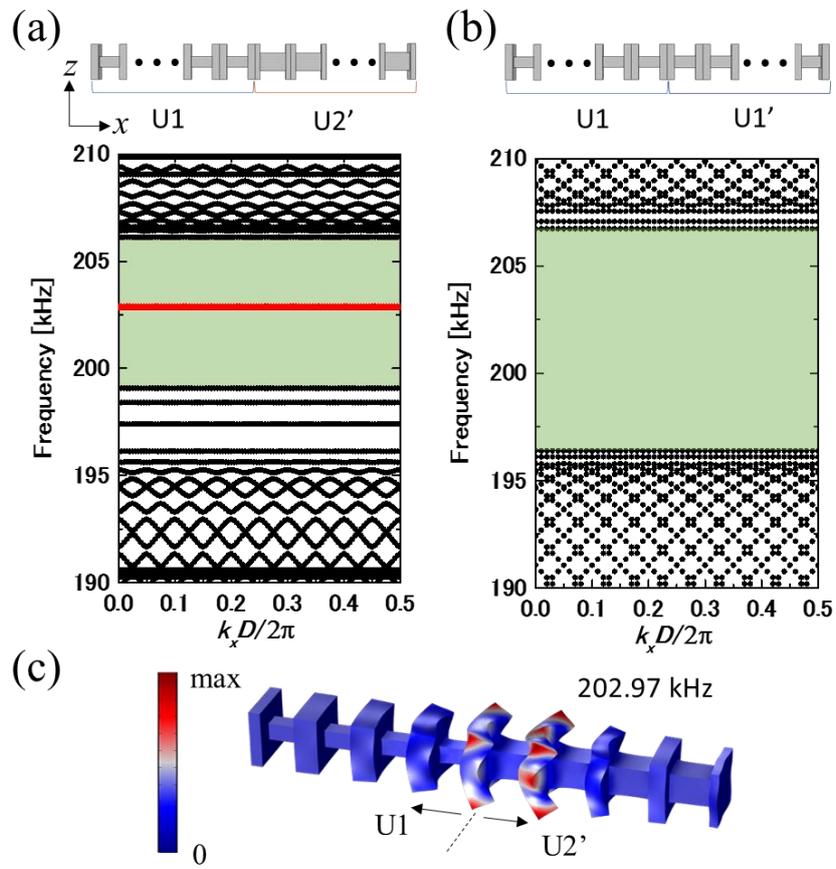

Fig. 2.

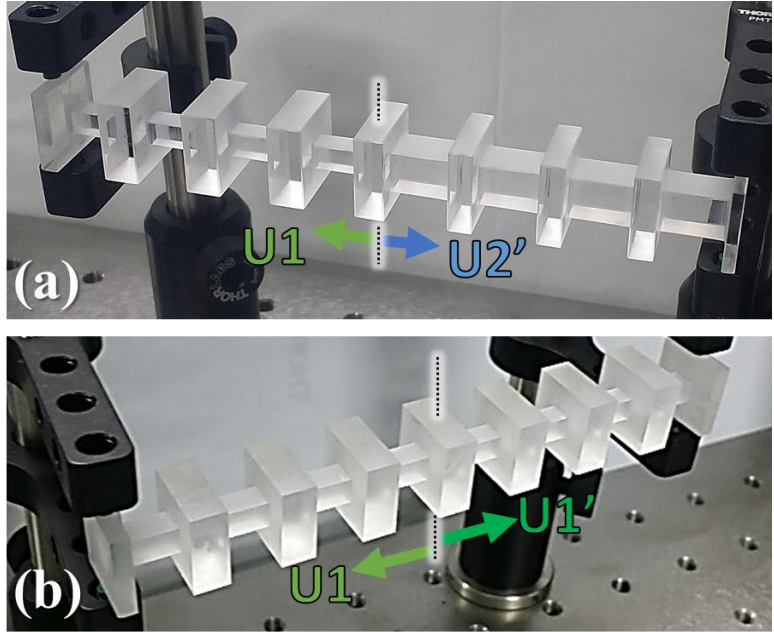

Fig. 3.

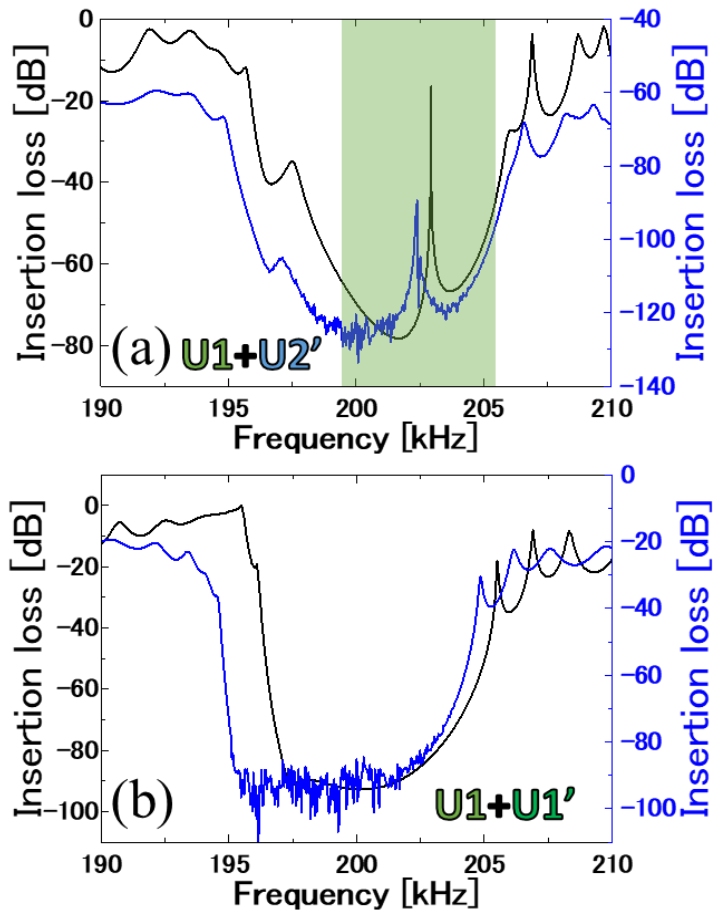
Fig. 4.

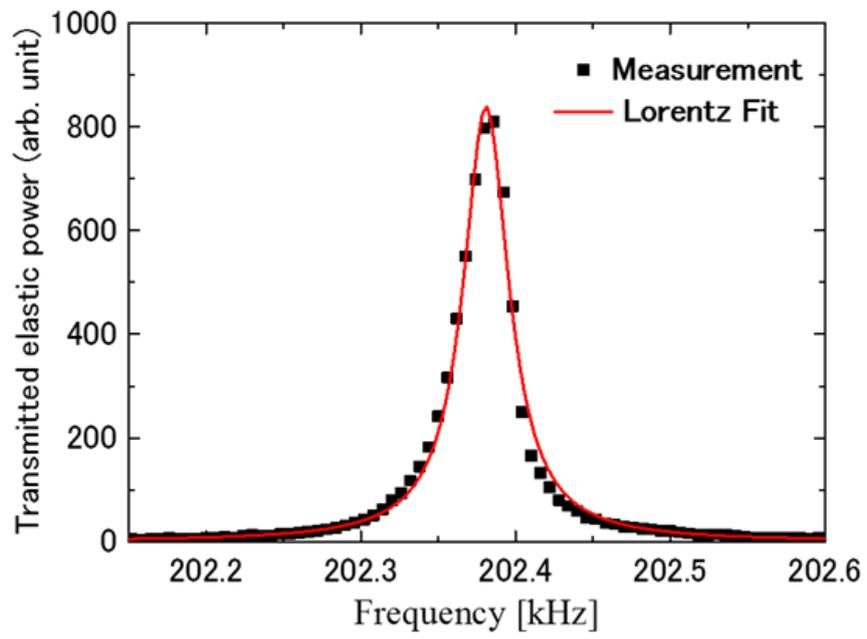

Fig. 5.

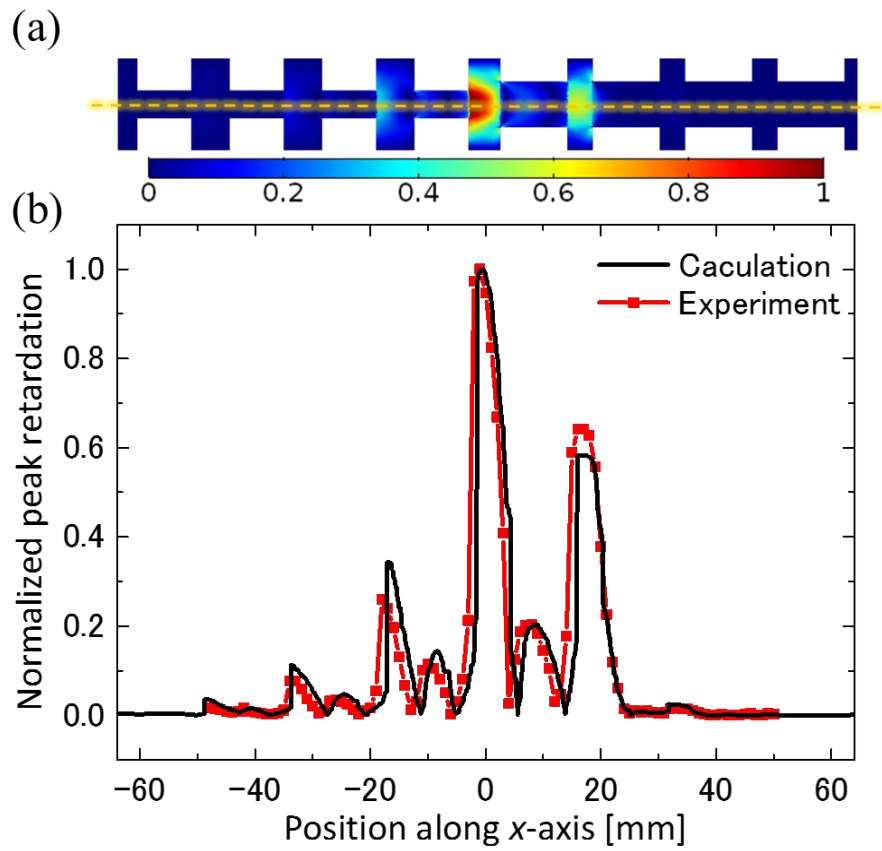

Fig. 6.

# Supplement data

## S1. Band structures of U1 and U2 with topological characteristics

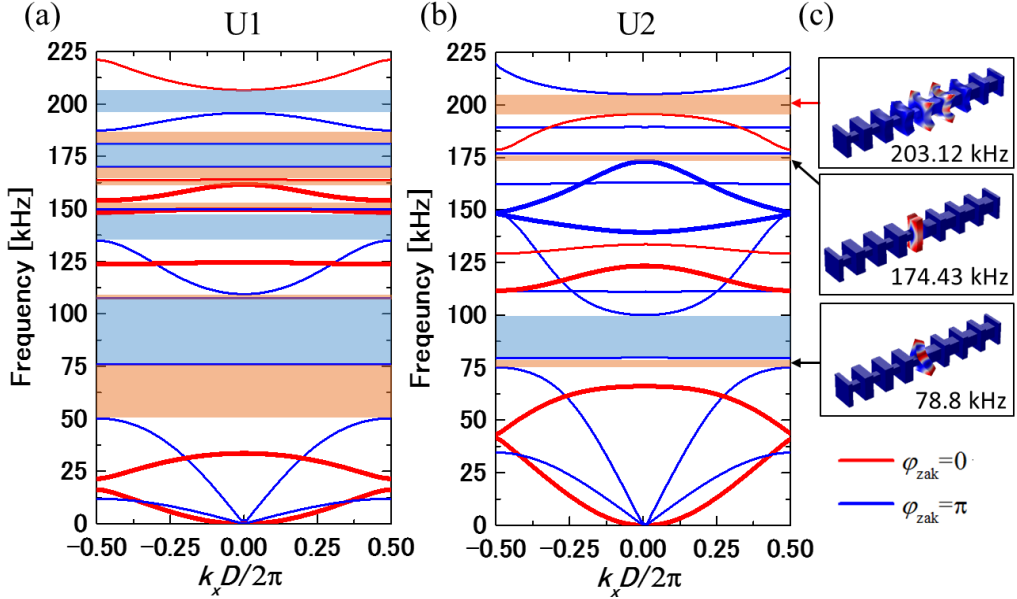

Fig. S1. Phononic band structure of U1 (a) and U2 (b) with Zak phase of each band. The values of Zak phase 0 ($\pi$) are shown in red (blue) color. Thick curves indicate doubly degenerate bands due to symmetrical cross section of the unit cell, which have same Zak phases each other. The color of PnBG region indicates different topological characteristics determined by summation of the Zak phase below the PnBG (c) Total displacement distributions of topological interface modes of a composite PnC consisting of U1 and U2 with $N=4$ for each component PnC. Eigenfrequency of each mode calculated by finite element method is indicated by arrows in overlapped PnBG region which are topologically distinct. The red arrow in the largest common PnBG is our target frequency of the topological interface state.

The topological property of the band gap depends on the summation of the Zak phases of all the bands below the gap whether the summation is either even or odd integer multiplied by $\pi$ when the system has inversion symmetry[1,2]. Figure S1 (a) and (b) shows the calculated phononic band structures with the Zak phase of each band of U1 and U2, respectively. The values of Zak phase 0 ($\pi$) are shown in red (blue) color. Several bands are doubly degenerate due to symmetric geometry of the unit cell in $yz$-plane as shown in the Fig. 1(a). The thick curves indicate such degenerate bands which have same Zak phases each other. Different topological property of the PnBG is indicated by different color of the PnBGs, either blue or orange. The topological interface states in the composite PnC, U1+U2, is guaranteed to exist in topologically different PnBGs due to the bulk-edge correspondence principle. There are three-overlapped PnBGs which are topologically distinct beween U1 and U2. The topological interfaces modes with total displacement fields in the composite PnC, U1+U2 consisting of four-unit cells for each components, are shown in Fig. S1(c). The interface states were found in the overlapped PnBGs which corresponding frequencies are indicated by arrows. The red arrow is our target frequency of topological interface state in the largest common PnBG with different topological properties, which provides strong localized state.

## S2. Topological phase transition in structural parameter space

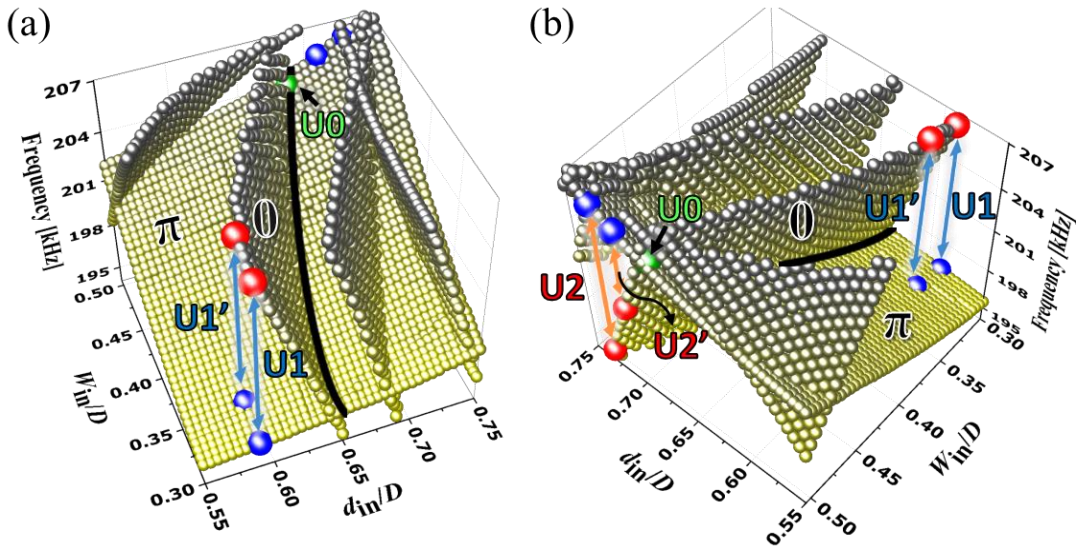

Fig. S2. The eigenfrequencies of eigenstates at $k_x=0$ near the PnBG near 200 kHz as a function of $W_{in}/D$ and $d_{in}/D$. (a) and (b) are a front and back view of the plot, respectively. Two surfaces labeled as $\pi$ and 0 are formed by the change of two band edge frequencies of the PnBG. The values of $\pi$ and 0 present the Zak phases of the bands including each band edge mode shown in blue and red circles, respectively. Four different configurations U1, U1', U2 and U2' discussed in the main text are plotted. Their PnBG widths are indicated by double-headed arrows. Different color of the arrows shows topologically distinct properties. The line of intersection between the two surfaces indicates a topological phase transition node (black curve). The topological phase transition point in the case of U0 is shown in green circle on the curve.

Figure S2 (a) and (b) show the eigenfrequencies of eigenstates near the PnBG near 200

kHz at $k_x=0$ as a function of $W_{in}/D$ and $d_{in}/D$, showing a front and back view of the graph, respectively. Two surfaces labeled as π and 0 in the parameter space were formed by the change of two band edge frequencies of the PnBG. The PnBG closes and reopens by tuning the values of $W_{in}/D$ and $d_{in}/D$, passing through a topological transition node shown in the black line of intersection between the two surfaces. Two configurations, U1' ($W_{in}/D = 0.33, d_{in}/D = 0.59$) and U2' ($W_{in}/D = 0.501, d_{in}/D = 0.72$), are also considered as components of composite PnCs as discussed in the main text. The PnBG width of each configuration is indicated by double-headed arrow which different color is to show that the PnBGs of U1 (or U1') and U2 (or U2') are topologically different. It is worth noting that the area of parameter space allowing topologically distinct complete PnBGs is limited since other surfaces formed by different mode frequencies at $k_x=0$ prevent the wide complete PnBGs from appearing at the same frequency ranges in different configuration. Therefore, the structure parameter should be carefully selected in order to achieve wide enough complete PnBGs which provide strongly-confined topological interface states.

## S3. Band structures of U1' and U2'

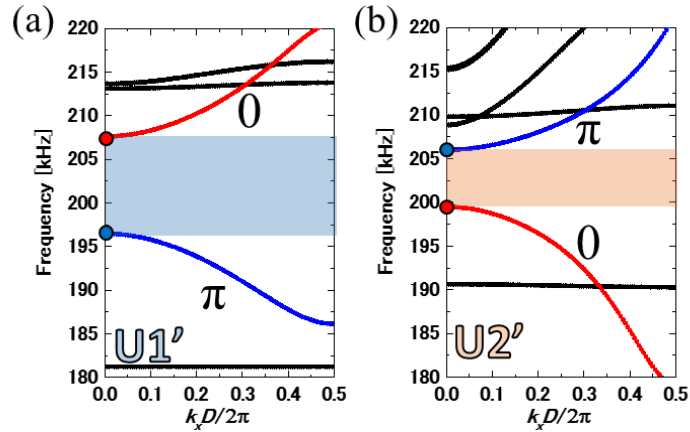

Fig. S3. Phononic band structures of U1' (a) and U2' (b), respectively. U1' and U2' have topologically distinct PnBGs each other. As discussed in Fig. 1 in the main text, the symmetries of the band-edge states are reversed in the case of U2' via band inversion process, corresponding to the switched colors of the circles at $k_x=0$ in (b).

Figure S3 (a) and (b) shows the phononic band structures of U1' and U2', respectively.

The values of π and 0 indicate the Zak phases of two bands near the PnBGs. The PnBG of each configuration with the topological characteristic corresponds to the double-headed arrow at its structural parameter in Fig. S2. As can be seen in Fig. S2 (b), the PnBG width of U1' is similar with that of U1, while U2' has relatively narrower PnBG than those of the other configurations. This induces flat bands below the common PnBG in the band structure using the supercell consisting of U1 and U2' as shown in Fig. 2(a). As similar with the relation between U1 and U2, U1' and U2' have distinct topological properties, while U1 (U2) and U1' (U2') are topologically identical.

## S4. Transmission spectra of U1, U1' and U2'

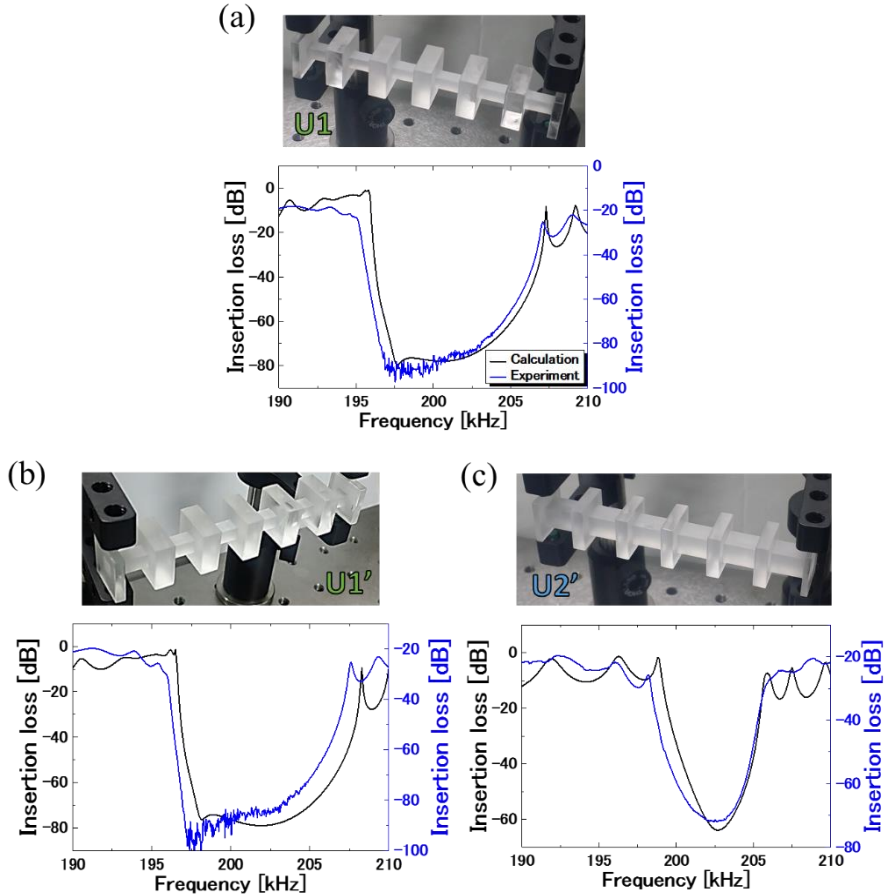

Fig. S4. Measured (blue) and calculated (black) transmission spectra with pictures of the measured samples, (a) U1, (b) U1' and (c) U2', respectively.

Figure S4 (a-c) shows transmission spectra of samples with U1, U1' and U2', respectively. All samples consists of 6-unit cells. The measured transmission spectra shown in blue are well matched with the calculation shown in black. Each frequency range where strong attenuation of elastic waves occurs shows a good agreement with the PnBG in the band structure as shown in Fig. 1(b) and Fig. S3. As predicted from the band structure of U2', its transmission spectrum shows relatively narrower frequency range where the PnBG effect occurs. However, there is still overlapped frequency range from 198.32 to 205.84 kHz with that of U1, where a single topological interface state is guaranteed to exist when the two PnCs are connected.

## S5. Methods to measure and calculate a spatial distribution of the topological interface states

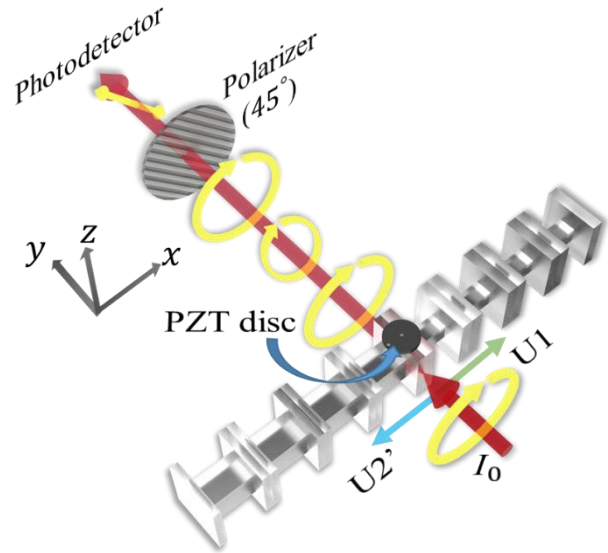

Fig. S5. Schematic of the experimental setup for measuring mode distribution of the interface states in a composite PnC consisting of U1 (*N*=4, left) and U2 (*N*=4, right).

The experiment setup used for the measurement of mode distribution is shown in Fig. S5. A PZT disc was attached at the interface to excite the topological elastic states. The PZT disc was driven by applying a 2.5-V peak-to-peak sinusoidal voltage at the interface mode frequency, $\omega_c/2\pi$ ~202.38 kHz. The induced birefringence through the photoelastic effect produces a time-varied phase retardation between *x*- and *z*-polarized components of the light passing through the composite PnC. A He–Ne laser (wavelength $\lambda_{\text{op}}$ =633 nm) was used as a light source. Transmitted light intensity $I_c$ through a polarizer along $45°$ with respect to the *x*-axis was measured by silicon photo detector coupled into an oscilloscope. In the case of circular-polarized incident light, the general expression for the measured intensity $I_c$ is expressed as

$$I_c = I_0[1 + 2J_1(A_0)\cos(\omega_c t) - 2J_3(A_0)\cos(3\omega_c t) + \cdots]/2, \qquad (S1)$$

where $A_0$ is the peak retardation. Using a spectrum analyzer, we measured the fundamental frequency component $J_1(A_0)$, which is proportional to $A_0$ when $A_0$ is small. By scanning the position of the incident light beam along the *x*-axis, a spatial distribution of the peak retardation was measured. We also calculated peak retardation $A_0(x,z)$ using material parameters of fused silica as same as our previous work[3]. The induced local birefringence distribution $\Delta n\,(x,z)$ at each *xz*-plane is not uniform along *y*-axis due to the quasi-longitudinal nature of the interface mode as shown in Fig. 5(b). Therefore, accumulated retardation distribution $A_0(x,z)$ after the light passes through the structure is expressed below as the integral of the induced birefringence over light propagation length *d* which also depends on the thickness of the structure along *y*-axis:

$$A_0(x,z) = \frac{2\pi}{\lambda_{op}} \int_0^d \Delta n\,(x,z)\,dy. \qquad (S2)$$

Since $\Delta n\,(x,z)$ is proportional to the difference between *x*- and *z*- component of strain tensor amplitude at each position, i.e. $\Delta n\,(x,z) \propto e_{xx} - e_{zz}$, the spatial distribution of peak retardation reflects the strain fields of the topological elastic waves.

[1] M. Xiao, Z.Q. Zhang, and C.T. Chan, Phys. Rev. X **4**, 021017 (2014).
[2] J. Zak, Phys. Rev. B **32**, 2218 (1985).
[3] I. Kim, S. Iwamoto, and Y. Arakawa, Jpn. J. Appl. Phys. **55**, 08RD02 (2016).